\documentclass[reprint, amsmath, amssymb, aps,superscriptaddress]{revtex4-1}
\usepackage{graphicx}
\usepackage{dcolumn}
\usepackage{bm}
\usepackage{mathrsfs}
\usepackage{tikz}
\usepackage{hyperref}
\input epsf.tex
\usepackage{amssymb}
\usepackage{amsmath}
\usepackage{amsthm}
\usepackage{amsopn}
\usepackage{ulem}
\usepackage{cancel}
\usepackage{color}
\usepackage{float}
\usepackage{comment}
\usepackage{multirow}

\hypersetup{
    colorlinks=true,       
    linkcolor=blue,          
    citecolor=blue,        
    filecolor=magenta,      
    urlcolor=blue           
}

\begin{document}

\title{Non-local interference in arrival time}
\author{Ali Ayatollah Rafsanjani}
\email{aliayat@physics.sharif.edu}
\affiliation{Department of Physics, Sharif University of Technology, Tehran, Iran}
\affiliation{School of Physics, Institute for Research in Fundamental Sciences (IPM), Tehran, Iran}

\author{MohammadJavad Kazemi}\email{kazemi.j.m@gmail.com}
\affiliation{Department of Physics, Faculty of Science, University of Qom, Qom, Iran}
\author{Vahid Hosseinzadeh}
\author{Mehdi Golshani}
\affiliation{School of Physics, Institute for Research in Fundamental Sciences (IPM), Tehran, Iran}

\begin{abstract}
Although position and time have different mathematical roles in quantum mechanics, with one being an operator and the other being a parameter, there is a space-time duality in quantum phenomena---a lot of quantum phenomena that were first observed in the spatial domain were later observed in the temporal domain as well. In this context, we propose a modified version of the double-double-slit experiment using entangled atom pairs to observe a non-local interference in the arrival time distribution,  which is analogous to the non-local interference observed in the arrival position distribution \cite{mahler2016experimental,strekalov1995observation}. However, computing the arrival time distribution in quantum mechanics is a challenging open problem \cite{vona2013does,rafsanjani2023can}, and so to overcome this problem we employ a Bohmian treatment. Based on this approach, we numerically demonstrate that there is a complementary relationship between the one-particle and two-particle interference visibilities in the arrival time distribution, which is analogous to the complementary relationship observed in the position distribution \cite{bergschneider2019experimental,Georgiev2021}. These results can be used to test the Bohmian arrival time distribution in a strict manner, i.e., where the semiclassical approximation breaks down. Moreover, our approach to investigating this experiment can be applied to a wide range of phenomena, and it seems that the predicted non-local temporal interference and associated complementary relationship are universal behaviors of entangled quantum systems that may manifest in various phenomena.

\end{abstract}

\maketitle

\section{Introduction}

In quantum theory, several effects that were initially observed in the spatial domain have subsequently been observed in the time domain. These effects include a wide range of phenomena such as diffraction in time \cite{moshinsky1952diffraction,tirole2023double,brukner1997diffraction,goussev2013diffraction}, interference in time \cite{szriftgiser1996atomic,ali2009quantum,kaneyasu2023time,rodriguez2023optical}, Anderson localization in time \cite{sacha2015anderson,sacha2016anderson} and several others \cite{hall2021temporal,coleman2013time,ryczkowski2016ghost,kuusela2017temporal,zhou2022space}. To extend this line of research, we propose a simple experimental setup that can be used to observe a non-local interference in \textit{arrival time}, which is analogous to the non-local interference in arrival position observed in entangled particle systems \cite{greenberger1993multiparticle,strekalov1995observation,hong1998two,kofler2012einstein,braverman2013proposal,kaur2020quantum,kazemi2023detection}.

\begin{figure}[t!]
	\centering
	\begin{tikzpicture}
		\draw (0,0) node[above right]{\includegraphics[width=1\linewidth, trim={0cm 0cm 0cm 0cm}]{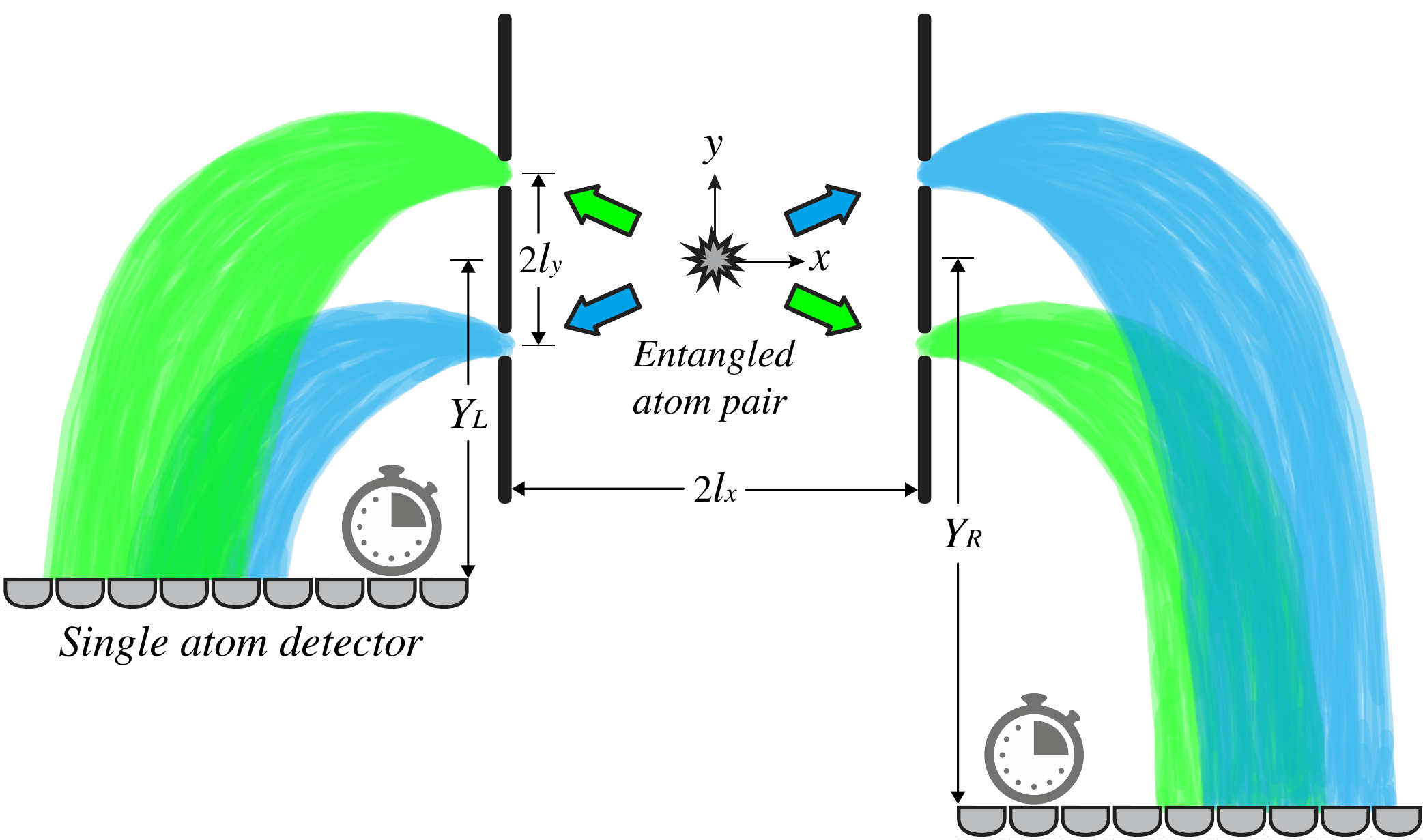}};
	\end{tikzpicture}
	\caption{\label{schematic} Schematic representation of the double-double-slit setup. The source emits pairs of entangled particles, each of which passes through a double-slit and then undergoes a free fall. Two arrays of fast particle detectors are placed on both sides, recording the detection events. $Y_L/Y_R$ represent the vertical distance from the origin to the left/right detection screen, $2l_y$ is the vertical distance between slits, and $2l_x$ is the horizontal distance between the slits.}
	\label{Setup-fig}
\end{figure}

The proposed experimental setup involves a double-double-slit arrangement in which a source emits pairs of entangled atoms toward slits \cite{gneiting2013nonlocal, kofler2012einstein}. Such entangled atoms can be produced, for example, via a four-wave mixing process in colliding Bose-Einstein condensates \cite{perrin2007observation, khakimov2016ghost}. As shown in Fig.\,\ref{Setup-fig}, the atoms fall due to the influence of gravity, and then they reach horizontal fast single-particle detectors, which record the arrival time and arrival position of the particles. In fact, a similar arrangement has previously been proposed for observing non-local two-particle interference in arrival \textit{position} distribution \cite{kofler2012einstein}. The critical difference between our setup and theirs is that the slits in our setup are not placed at the same height. This leads to height-separated wave packets that spread in space during falling and overlap each other. Moreover, we do not consider the horizontal screens at the same height, so the particles may be detected at completely different timescales. Our study indicates that these apparently small differences lead to significant interference in the two-particle arrival time distribution, which did not exist in the previous versions of the experiment. This phenomenon is experimentally observable, thanks to the current single-atom detection technology. Our numerical study shows that the required space-time resolution in particle detection is achievable using current single-atom detectors, such as the recent delay-line detectors described in \cite{Keller2014Bose,khakimov2016ghost} or the detector used in \cite{kurtsiefer19962A,kurtsiefer1997measurement}.

The theoretical analysis of the proposed experiment is more complex than that of the conventional double-double-slit experiment due to at least two reasons. Firstly, since the two particles are not observed simultaneously, the wave function of the two particles collapses to a single-particle wave function at the time when the first particle is detected in the middle of the experiment. Secondly, the theoretical analysis of arrival time distribution is more complex than that of arrival position distribution. This is because, in the mathematical framework of orthodox quantum mechanics, position is represented by a self-adjoint operator, while time is just treated as a parameter \footnote{In fact, Pauli showed that there is no self-adjoint time operator canonically conjugate to the Hamiltonian if the Hamiltonian spectrum is discrete or has a lower bound \cite{Pauli1958Encyclopedia}.}. As a result, the Born rule cannot be directly applied to the time measurements. This fact, coupled with other issues such as the quantum Zeno paradox \cite{misra1977zeno,porras2014quantum}, leads to some ambiguities in calculating the arrival time distribution \cite{Allcock1969,mielnik1994screen,leavens2002standard,vona2013does,sombillo2016particle,das2021times,das2021questioning}. In fact, there is no agreed-upon method for calculating the arrival time distribution, although several different proposals have been put forth based on various interpretations of quantum theory \cite{grot1996time,leavens1998time,halliwell1998decoherent,marchewka2002measurement,galapon2004confined,nitta2008time,anastopoulos2012time,Maccone2020,roncallo2023does,tumulka2022detection,rafsanjani2023can}. 

Most of the arrival time distribution proposals are limited to simple cases, such as a free one-particle in one dimension, and are not yet fully extended to more complex situations, such as our double-double-slit setup. Nevertheless, the Bohmian treatment seems suitable for analyzing the proposed setup since it can be unambiguously generalized for multi-particle systems in the presence of external potentials. Thus, in this paper, we investigate the proposed experiment using the recent developments in the Bohmian arrival time for entangled particle systems, including detector back-effect \cite{tumulka2022detection,kazemi2023detection}. The results could contribute to a better understanding of the non-local nature of quantum mechanics in the time domain. Moreover, beyond the proposed setup, our theoretical approach has potential applications in related fields such as atomic ghost imaging \cite{khakimov2016ghost,hodgman2019higher},  quantum test of the weak equivalence principle with entangled atoms \cite{geiger2018proposal}, and  state tomography via time-of-flight measurements \cite{brown2023time,kurtsiefer1997measurement,das2022double}.

This paper is organized as follows. We present the theoretical framework in Sec.\,\ref{Theoretical framework}. We then discuss the numerical results and the physical insights derived from them in Sec.\,\ref{Results}, including the signal locality, the complementarity between one-particle and two-particle interference visibilities, and the screen back-effect. In Sec.\,\ref{semiclassical-section}, we compare the Bohmian approach with the semiclassical approximation. We conclude with a summary and an outlook in Sec.\,\ref{Summary}. 
 
\section{Theoretical framework}\label{Theoretical framework}

Bohmian mechanics, also known as pilot wave theory, is a coherent realistic version of quantum
theory, which avoids the measurement problem \cite{bell1990against,goldstein1998quantum}. In the Bohmian interpretation, in contrast to the orthodox interpretation, the wave function does not give a complete description of the quantum system. Instead, the actual trajectories of particles are taken into account as well, and this can provide a more intuitive picture of quantum phenomena \cite{benseny2014applied}. Nonetheless, it has been proved that in the quantum equilibrium condition \cite{durr1992quantum,valentini2005dynamical}, Bohmian mechanics is experimentally equivalent to orthodox quantum mechanics \cite{bohm1952suggested,durr2004quantum} \textit{insofar as the latter is unambiguous} \cite{bell1980broglie,Das2019Arrival,ivanov2017exit}; e.g., in usual position or momentum measurements at a specific time. In recent years, Bohmian mechanics has gained renewed interest for various reasons \cite{albareda2014correlated,larder2019fast,xiao2019observing,foo2022relativistic}. One of these reasons is the fact that Bohmian trajectories can lead to clear predictions for quantum characteristic times, such as tunneling time duration \cite{zimmermann2016tunneling,douguet2018dynamics} and arrival time distribution \cite{leavens1998time,Das2019Arrival,kazemi2023detection}. 

Here, we investigate the proposed double-double slit setup using Bohmian tools. According to Bohmian Mechanics, the state of a two-particle system is determined by the wave function $\Psi(\bm r_1,\bm r_2)$ and the particles' actual positions $(\bm R_1,\bm R_2)$. 
The time evolution of the wave function is given by a two-particle Schrödinger equation
\begin{equation}\label{Schrodinger}
i\hbar\frac{\partial}{\partial t}\Psi_t(\bm r_1,\bm r_2)=\sum_{i=1,2}\frac{\hbar^2}{2m_i}\nabla^2_i\Psi_t+V_i(\bm r_i)\Psi_t,
\end{equation}
which in the proposed setup 
$V_i(\bm r_i)=-m_i\bm g.\bm r_i$ and $\bm g$ represents the gravitational field.  
The particles dynamics are given by two first-order differential equations in configuration space, the ``guidance equations",  
\begin{align}\label{guiding}
\frac{d}{dt}\bm R_i(t)=\bm v_i^{\Psi_t}(\bm R_1(t),\bm R_2(t)),
\end{align} 
where $i=1,2$ and $\bm v_i^{\Psi_t}$ are the velocity fields associated with the wave function $\Psi_t$; i.e. 
$\bm v_i^{\Psi_t}{=}(\hbar/m_i)\Im(\nabla_i\Psi_t/\Psi_t)$ \cite{durr1995quantum}.
When the particle $1$, for example, is detected at time $t=t_c$, the two-particle wave function 
collapses \textit{effectively} to a one-particle wave function, i.e. as $\Psi_{t_c}(\bm r_1,\bm r_2)\to\psi_{t_c}(\bm r_2)$, where \cite{Durr2004,Durr2010On},
\begin{equation}\label{collapsed wave function}
\psi_{t_c}(\bm r_2)=\Psi_{t_c}(\bm R_1(t_c),\bm r_2),
\end{equation}
which is known as the “conditional wave function” in Bohmian formalism \cite{durr1995quantum,Norsen2014}. For $t>t_c$, the time evolution of the wave function is given by following the one-particle Schrödinger equation
\begin{equation}\label{one particle Schrodinger}
i\hbar\frac{\partial}{\partial t}\psi_t(\bm r_2)=\frac{-\hbar^2}{2m_2}\nabla^2_2\psi_t(\bm r_2)+V_2(\bm r_2)\psi_t(\bm r_2),
\end{equation}
and the remaining particle motion is determined by the associated one-particle guidance equation,
\begin{equation}\label{one particle guiding}
\frac{d}{dt}\bm R_2(t)=\bm v_2^{\psi_t}(\bm R_2(t)),
\end{equation}
where $\bm v_2^{\psi_t}{=}(\hbar/m_2)\Im(\nabla_2\psi_t/\psi_t)$.
It is important to note that, in general, a conditional wave function does not obey the Schrodinger equation \cite{durr2012quantum}. However, in a measurement situation, the interaction of the detected particle with the environment (including the detection screen) cancels any entanglement between undetected and detected particles,
 due to the decoherence process \cite{rovelli2022preparation}. Therefore, in this situation, after the measurement process, the conditional wave function represents the ``effective wave function" of the undetected particle \cite{durr2012quantum,teufel2009bohmian}, which satisfies the one-particle Schrodinger equation \cite{durr2012quantum,tumulka2022detection}.
 
We focus our study on the propagation of the wave function from the slits to the detection screens. Thus, one can consider the initial wave function as follows \cite{Braverman2013,Georgiev2021,pathania2022characterization}:
\begin{eqnarray}\label{}
\Psi_{t_0}(\bm r_1,\bm r_2)\!=\!N\!\left[(\frac{1-\eta}{2})\Psi_{t_0}^{(\times)}(\bm r_1,\bm r_2)\!+\!(\frac{1+\eta}{2})\Psi_{t_0}^{(||)}(\bm r_1,\bm r_2)\right],\nonumber
\end{eqnarray}
in which $N$ is a normalization constant,
\begin{eqnarray}\label{Initial wave function}
\Psi_{t_0}^{(\times)}(\bm r_1,\bm r_2)&=&[g_{u}^+(\bm r_1)g_{d}^-(\bm r_2)+g_{d}^+(\bm r_1)g_{u}^-(\bm r_2)]+ 1\leftrightarrow 2,\nonumber\\
\Psi_{t_0}^{(||)}(\bm r_1,\bm r_2)&=&[g_{u}^+(\bm r_1)g_{u}^-(\bm r_2)+g_{d}^+(\bm r_1)g_{d}^-(\bm r_2)] + 1\leftrightarrow 2,\nonumber
\end{eqnarray}
and
\begin{eqnarray}\label{fGG-def}
g_{u}^\pm(x,y)&=&G(x;\sigma_x,\pm l_x, \pm u_x)G(y;\sigma_y, +l_y,  +u_y),\nonumber\\
g_{d}^\pm(x,y)&=&G(x;\sigma_x,\pm l_x, \pm u_x)G(y;\sigma_y, -l_y,  -u_y),\nonumber
\end{eqnarray}
 \begin{figure}[h!]
	\centering
	\begin{tikzpicture}
		\draw (0,0) node[above right]{\includegraphics[width=1\linewidth, trim={0.5cm 0cm -0.1cm 0cm}]{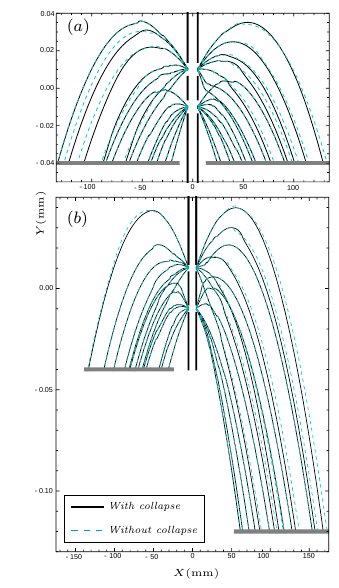}};
	\end{tikzpicture}
	\caption{Collapse effect on the trajectories of entangled helium atom pairs for $\eta=-1$. The cyan curves show the trajectories without collapse, and the black curves show how they change due to the collapse effect. In panel (a), the screens are at the same height ($Y_L=Y_R=0.04$ mm), while in panel (b), they are at different heights ($Y_L=0.04$ mm and $Y_R=0.12$ mm).}
	\label{schematicTraj}
\end{figure}
where $G$ is a Gaussian wave function
$$G(x;\sigma, l,  u)=Ne^{-(x-l)^2/4 \sigma^2+ i m u (x-l)/\hbar}.$$
The Gaussian-type initial wave function is a minimal model which is commonly used in the literature (e.g., see \cite{Georgiev2021,pathania2022characterization,peled2020double,kaur2020quantum,sanz2023young,braverman2013proposal,guay2003two,golshani2001bohmian,nitta2008time}).
The wave function is symmetrized, as we have considered the particles as indistinguishable bosons.
The parameter $\eta$ controls the degree of entanglement of the initial state. It is easy to see that  $\eta=0$ leads to a separable state, whereas $|\eta|=1$ leads to a maximally entangled state; for
$\eta=+1$ the state is maximally correlated, and for $\eta=-1$ the state is maximally anticorrelated \cite{Georgiev2021}. For more details, in panel (a) of Fig.\,\ref{visibility}, the entanglement entropy is plotted as a function of $\eta$, which is a measure of the degree of quantum entanglement between two particles and is defined as $-\text{Tr}(\hat\rho\text{Log}\hat\rho)$, where $\hat\rho$ is the one-particle reduced density matrix \cite{srednicki1993entropy,horodecki2009quantum}. It is worth noting that even without using the slits, this kind of initial wave function could be produced and reliably controlled using optical manipulation \cite{goussev2015manipulating,akbari2022optical,bergschneider2019experimental} of an entangled state generated from colliding Bose-Einstein condensates \cite{perrin2007observation,jaskula2010sub,khakimov2016ghost}.
Furthermore, since the free two-particle Hamiltonian is separable, the time evolution of this wave function can be found from Eq.\,\eqref{Schrodinger} analytically as
 \begin{eqnarray}\label{}
\Psi_{t}(\bm r_1,\bm r_2)\!=\!N\!\left[(\frac{1-\eta}{2})\Psi_{t}^{(\times)}(\bm r_1,\bm r_2)\!+\!(\frac{1+\eta}{2})\Psi_{t}^{(||)}(\bm r_1,\bm r_2)\right]\nonumber
\end{eqnarray}
in which functions $\Psi_{t}^{(\times)}(\bm r_1,\bm r_2)$ and $\Psi_{t}^{(||)}(\bm r_1,\bm r_2)$ are constructed out of time dependent Gaussian wave functions $G_t$ as 
\begin{eqnarray} \label{Gaussian wave}
    G_t(x;\sigma, l,  u, a) &=& (2 \pi s_t^2)^{-\frac14} e^{i \frac{ma}{\hbar}l t}e^{[\frac{-(x-l-u t-a t^2/2)^2}{4\sigma s_t}]}\nonumber \\    
    &\times & e^{i\frac{m}{\hbar}[(u-at).(x-l-\frac{u t}{2})-a^2 t^3/6]} 
\end{eqnarray}
where $a$ represents the acceleration component and $s_t = \sigma (1+i  \frac{t\hbar}{2 m \sigma^2})$ \footnote{The Gaussian solution of the Schrödinger equation for a particle under uniform force was first introduced by de Broglie \cite{de1930introductio} and rephrased by Holland in \cite{holland1995quantum}, but according to our investigation, none of them satisfy the Schrödinger's equation exactly: To satisfy the wave equation, an additional time-dependent phase, i.e. our first exponential term in \eqref{Gaussian wave}, is needed.}.
Using this wave function, the detection time and position of the first observed
particle are uniquely determined by solving Eq.\,\eqref{guiding}. Then, using Eq.\,\eqref{one particle guiding}, we can find the trajectories of the remaining particles after the first particle detection. 
\begin{figure*} 
\includegraphics[width=1.0\linewidth]{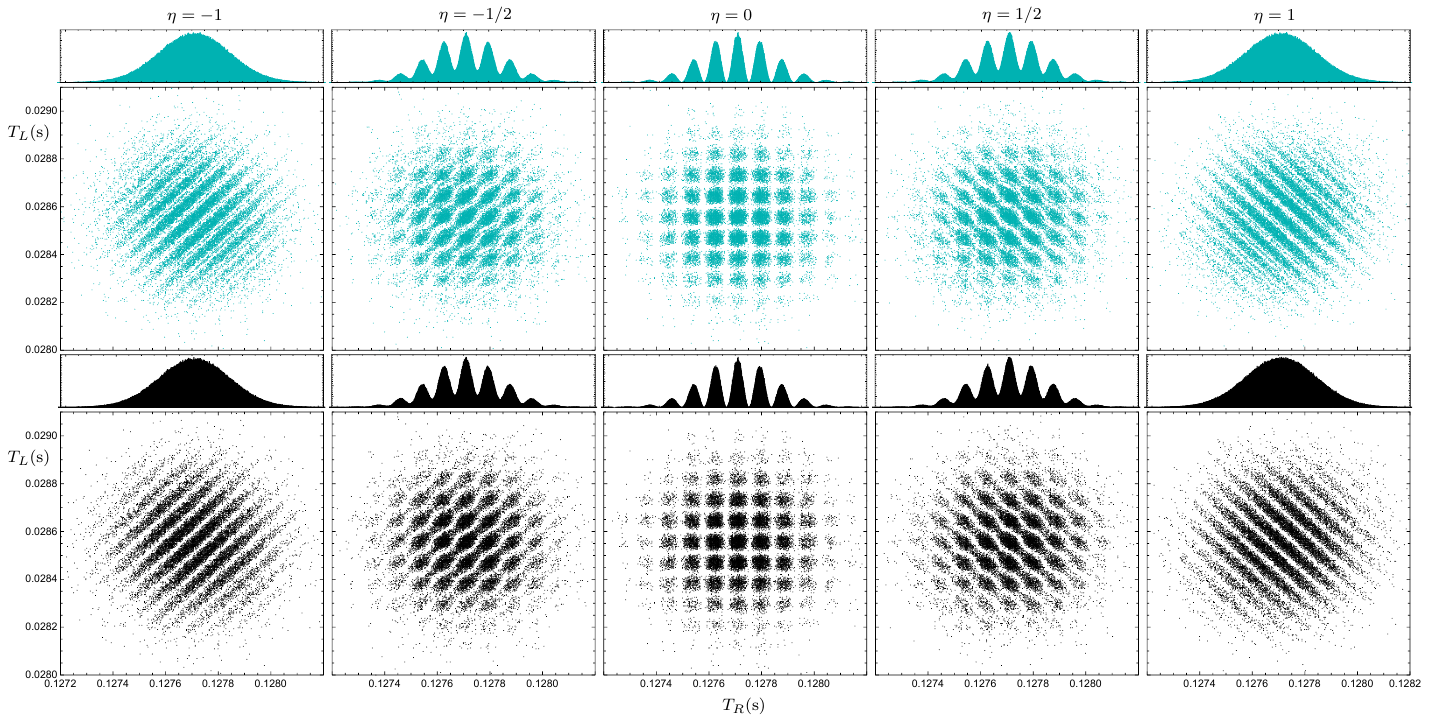}
\caption{One- and two-particle interference patterns of sodium atoms for different entanglement levels $\eta$, with and without the collapse effect. The scatter plots show the joint distributions of arrival times to the horizontal screens, with the dark-cyan plots ignoring the collapse effect and the black plots considering it. The left and right screens are placed at   $Y_L=4$ mm and $Y_R= 8$ cm, respectively. The histograms in each panel show the marginal distributions of arrival times to the right screen.}\label{EthaEffect}
\end{figure*}

Using trajectories, we can find the joint detection data distribution in $(t_L,x_L;t_R,x_R)$ space where $t_{L/R}$ is the detection time, and $x_{L/R}$ is the detection position on the left/right screen.  The probability density behind this distribution can be formally written as
\begin{align*}
P(t_L,x_L;t_R,x_R)=\int d \bm R^0\ |\Psi_0(\bm R^0)|^2 \hspace{3cm}\\
\times \prod_{i=L,R}\delta(t_i-T_i(\bm R^0))\delta(x_i-X_i{(\bm R^0)}),
\end{align*}
where  $T_{L,R}{(\bm  R^0_1, \bm  R^0_2)}$ and  $X_{L,R}{(\bm  R^0_1, \bm  R^0_2)}$ are the arrival time and position of the particle with initial condition $(\bm  R^0_1, \bm  R^0_2)$ to the left and right screen, respectively. Note, how the above joint distribution and, therefore, any marginal distribution out of it, is sensitive to the Bohmian dynamics through functions $T$ and $X$ and also to the Born rule by $|\Psi_0(\bm R^0)|^2$. The joint two-particle arrival time probability density is then defined as, 
\begin{align}
	\Pi(t_L, t_R) = \int\int P(t_L, x_L; t_R, x_R) dx_L dx_R.
\end{align}
The right and left marginal arrival time probability densities are also defined correspondingly as, 
\begin{align*}
	\Pi_{L}(t_L) = \int P(t_L, t_R) dt_R, \\
	\Pi_{R}(t_R) = \int P(t_L, t_R) dt_L.
\end{align*}
In a practical manner, the trajectories of the particles and the resulting arrival time distributions are numerically computed for an ensemble of particles whose initial positions are sampled from $|\Psi_0|^2$, and the corresponding results are described in the next section. 

\section{Results and Discussions}\label{Results}
In the numerical studies of this work, the parameters of the proposed setup have been chosen as $l_x \!=\! 5\times10^{-3}$ m and $l_y \!=\!10^{-5}$ m. Moreover, the initial wave packets widths and velocities are fixed at $\sigma_x\!=\!\sigma_y\!=\!10^{-6}$ m, $u_x=20$ m/s and $u_y\!=\!0$, respectively. These values are consistent with the proposed setup in reference \cite{Kofler2012}, in which colliding helium-4 atoms have been considered for producing an initial entangled state \cite{Perrin2007}. However, we also consider heavier atom pairs, which lead to a more visible interference pattern for some values of parameters and the locations of the screens.

In Fig.\,\ref{schematicTraj}, some of the Bohmian trajectories are plotted, for maximally anti-correlated helium atom pairs. In this figure, the cyan trajectories are without considering the collapse effect, and the black ones are with it. One can see that some of the black trajectories start to deviate from the cyan ones as the screens detect the counterpart particles and the conditional wave function now guides undetected particles. It is worth noticing that, the ensemble of trajectories can be experimentally reconstructed using weak measurement techniques \cite{kocsis2011observing,braverman2013proposal,mahler2016experimental}, which can be used as a test of this result.

\subsection{Complementary relation of visibilities}
In Figs.\,\ref{EthaEffect},  the joint arrival time distribution $\Pi(t_L, t_R)$ and the right marginal distribution $\Pi_{R}(t_R)$ are plotted, for sodium atom pairs in two cases: with collapse effect in black and without it in dark-cyan. In this figure, we see the one-particle and two-particle temporal interference patterns for fixed screen locations ($Y_L=4$ mm and $Y_R=80$ mm) and different values of the entanglement parameter $\eta$. The marginal distributions are generated using $10^6$ particles' trajectories, however, for clarity, only $10^4$ points are shown in the joint scatter plots. As previously mentioned, the maximum entanglement occurs when $|\eta|=1$; and when $\eta=0$, the particles are entirely uncorrelated. As one can see in Fig.\,\ref{EthaEffect}, the visibilities of the joint and marginal distributions have an inverse relation; when the one-particle interference visibility is maximal, the two-particle interference visibility is minimal, and vice versa.  This behavior represents a \textit{temporal} counterpart to the complementarity between the one-particle and two-particle interference visibilities of the arrival \textit{position} pattern, which can be observed in a double-double-slit configuration \cite{Georgiev2021, peled2020double}. In fact, an analogous behavior has been observed in the momentum distribution of entangled atom pairs \cite{bergschneider2019experimental}. However, it is important to remark that, the ‘‘time of flight measurement’’ technique, which is usually used to measure the momentum distribution in the context of cold-atom experiments, is a position measurement after a specific large time, not an arrival time measurement at a specific position \cite{vona2013does}.

A quantitative study of such complementary relationship is firstly discussed in the pioneering works of  Jaeger, Horne, Shimony, and Vaidman for discrete systems \cite{jaeger1993complementarity,jaeger1995two}.
In these works, it is shown that there is a trade-off between these two visibilities, such that their squares add up to one or less, $W^2+V^2\leq 1$, where $V$ and $W$ are one-particle and two-particles interference visibility, respectively \cite{jaeger1993complementarity}. Recently, this complementary relationship has been studied for some continuous-variables, i.e., position and momentum, in a double-double-slit configuration \cite{Georgiev2021}. Here we numerically study this complementary relation for arrival time interference patterns. In this regard, the visibilities associated with arrival time interference patterns, which are represented in Fig.\,\ref{EthaEffect}, are shown in Fig.\,\ref{visibility}---for more details of the visibility estimation method, see appendix \ref{AppendixB}.  As one can see in Fig.\,\ref{visibility}, the one-particle and two-particle visibilities have opposite behavior and the sum of their squared values satisfies the mentioned complementary relation.

\begin{figure}[h!]
\includegraphics[width= \linewidth]{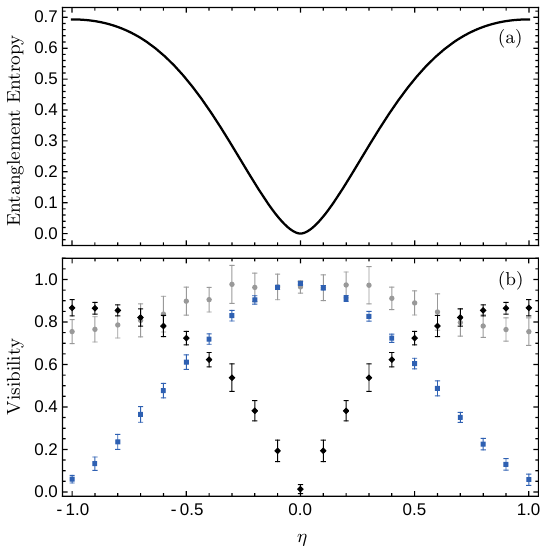}
\caption{Entanglement entropy and visibilities of one- and two-particle interference patterns of arrival time, as functions of $\eta$. Panel (a) show the entanglement entropy calculated from the initial wave function given by Eq.\,\eqref{Initial wave function}. Panel (b) shows the one-particle visibility $V$, the two-particle visibility $W$, and their squared sum $V^2+W^2$, represented by blue square, black diamond, and gray circle markers, respectively. The setup parameters and initial conditions are the same as in Fig.\,\ref{EthaEffect}, and the error bars are estimated from the counting statistics.}\label{visibility}
\end{figure}

\begin{figure*}[ht!]
\includegraphics[width=1.0\linewidth]{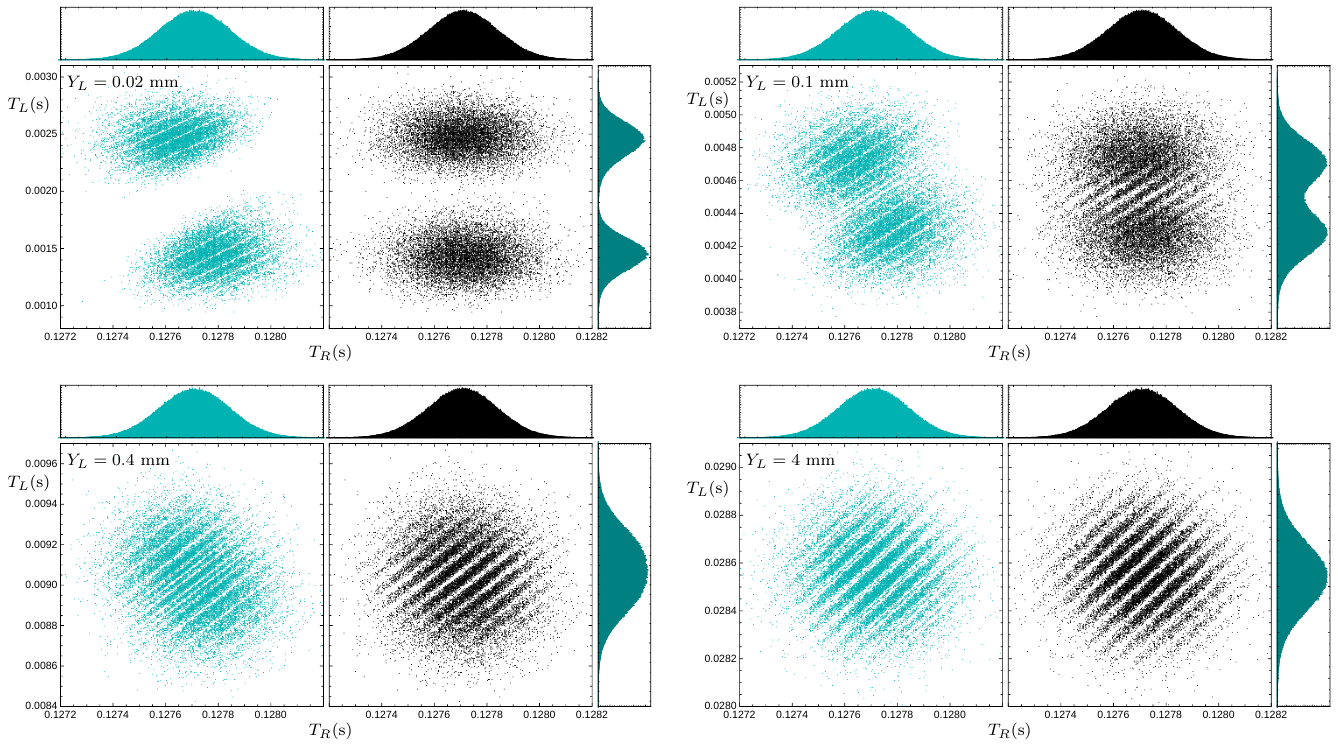}
\caption{Arrival time distributions of sodium atom pairs for different distances of the left screen. The left and right scatter plots in each panel show the joint distributions of arrival times to the horizontal screens for the cases without (dark cyan) and with (black) collapse, respectively. The upper and right histograms in each panel show the marginal distributions of arrival times to the right and left screens, respectively. The right screen is fixed at $Y_R=8$ cm, and $\eta=-1$. }\label{CollapseEffect}
\end{figure*}

\subsection{Collapse effect and signal locality}
As one can see in Fig.\,\ref {EthaEffect}, the correction of the two-particle arrival time distributions due to the collapse effect decreases with the turning off of the entanglement, and in $\eta=0$, interference patterns with and without correction are the same. In fact, in this case, we have $\Pi(t_L, t_R)=\Pi_{R}(t_R)\Pi_{L}(t_L)$. In Fig.\,\ref {CollapseEffect}, the one-particle and two-particle temporal interference patterns for different positions of the left screen are depicted, while the entanglement parameter is fixed to $\eta=-1$. The difference between patterns is obvious in the cases without collapse effect consideration (dark-cyan plots) and by including the effect (black plots). The closer the left screen is to the slits, the earlier the wave function reduction occurs, and its effect is more visible on the joint distribution. Note that, despite the fact that the collapse effect changes particles' trajectories and resulting joint distribution, this effect does not change the one-particle distribution patterns. This shows the establishment of the no-signaling condition, despite the manifest non-local Bohmian dynamics: The right marginal arrival time distribution, as a local observable quantity, turns out to be independent of whether there is any screen on the left or not, and if there is any, it is not sensitive to the location of that detection screen.
Note that this fact is not trivial because the well-known no-signaling theorem is proved for observable distributions provided by a POVM. However, in the general case, the \textit{intrinsic} Bohmian arrival time distribution cannot be described by a POVM, at least when the detector back-effect is ignored \cite{vona2013does,Das2019Exotic}. In the next subsection, we discuss more on the detector back-effect. 

\begin{figure*}[ht!]
	\centering
	\begin{tikzpicture}
		\draw (-40 ,0) node[above right]{\includegraphics[width=1\linewidth, trim={0cm 0cm 1.5cm 0cm}]{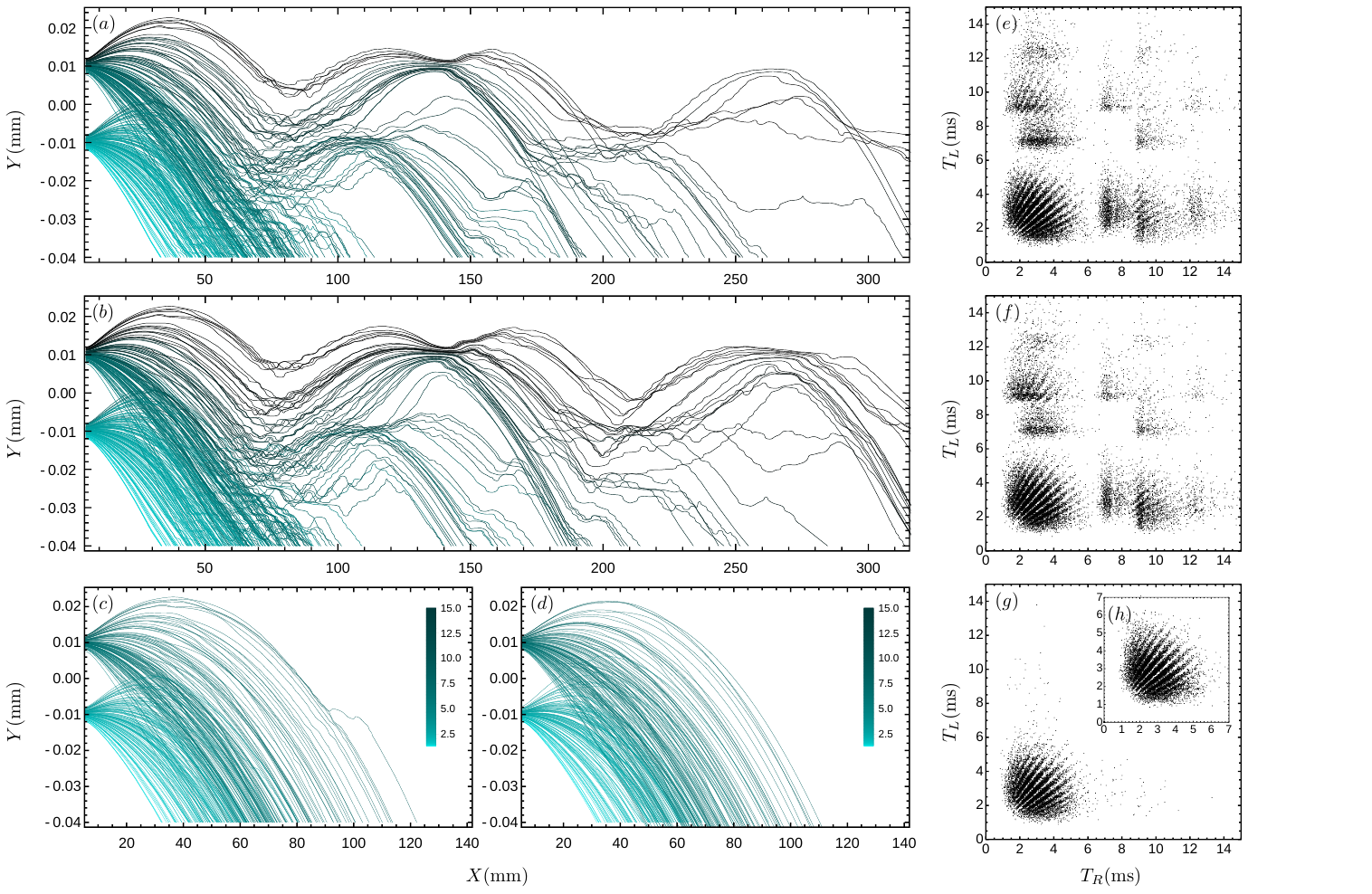}};
	\end{tikzpicture}
	\caption{Bohmian trajectories and joint arrival time distributions of helium atoms in the two double-slit setup, with different detector characterizing constants $\kappa$. The detectors are placed at $Y_R=Y_L =40\, \mu$m, and is considered as an absorbing boundary. The slit distances are $2l_y=20 \ \mu$m, the initial wave packets’ dispersions are $\sigma_x=\sigma_y=1 \  \mu$m, and the initial velocities are $u_x=20 $ m$/$s and  $u_y=0$. Panels (a), (b), and (c) show the trajectories for $\kappa=3\kappa_0$, $\kappa=\kappa_0/3$, and $\kappa=\kappa_0$, respectively. Panel (d) shows the trajectories without the detector back-effect, which are only truncated at the detector position. The colors of all trajectories are based on their arrival times on the screen and are labeled by color-bar in the unit of ms. Panels (e), (f), (g), and (h) show the joint arrival time distributions corresponding to the trajectories in panels (a), (b), (c), and (d), respectively.}
	\label{abctraj}
\end{figure*} 

\subsection{Detector back-effect}
The arrival distributions computed so far should be called \textit{ideal} or \textit{intrinsic} distributions \cite{Das2019Exotic}, since the influence of the detector, before particles detection, has been ignored in our theoretical manner. Such an idealization is commonly used in most previous studies of Bohmian arrival time distribution (for example, see \cite{Leavens_BohmianTOA,ali2009quantum,AliPhysRevA,MousaviGolshani,Das2019Exotic,Das2019Arrival}), and seems more or less to be satisfactory in many applications including the double-slit experiment \cite{das2022double,coffey2011reconstruction,kocsis2011observing}. Nonetheless, in principle, the presence of the detector could modify the wave function evolution, even before the particle detection \cite{mielnik2011time}. This is called the detector back-effect. To have a more thorough investigation of detection statistics, we should consider this effect. However, due to some fundamental problems, such as the measurement problem and the quantum Zeno effect \cite{misra1977zeno}, a complete investigation of the detector effects is problematic at the fundamental level, and it is less obvious how to model an ideal detector \cite{Allcock1969,mielnik1994screen,mielnik2011time}. 
 Nonetheless, some phenomenological non-equivalent models are proposed, such as the generalized Feynman path integral approach in the presence of an absorbing boundary \cite{Marchewka1998Feynman,Marchewka2000Path,Marchewka2001Survival,Marchewka2002}, the Schrödinger equation with a complex potential \cite{tumulka2022absorbing}, the Schrödinger equation with absorbing (or complex Robin) boundary condition \cite{werner1987arrival,Tumulka2022Distribution,tumulka2022detection,Dubey2021Quantum,tumulka2022absorbing}, and so on. In this section, we merely consider the absorbing boundary rule (ABR), which is compatible with the Bohmian picture and recently developed for multi-entangled particle systems \cite{tumulka2022detection}. The results of other approaches may not be the same \cite{rafsanjani2023can}---See also Section \ref{semiclassical-section}. So a detailed study of the differences is an interesting topic, which is left for future works.

 \textit{Absorbing Boundary Rule}---According to the ABR, the particle wave function $\psi$ evolves according to the free Schr\"{o}dinger equation, while the presence of a detection screen is modeled by imposing the following boundary conditions on the detection screen, $\bm{r}\in \mathbb{S}$, 
\begin{equation}\label{ABC_eq}
\bm n\cdot\nabla\psi=i\kappa\psi,
\end{equation}
where $\kappa \!>\!0$ is a constant characterizing the type of detector, in which $\hbar \kappa/m$ represents the momentum that the detector is most sensitive to. This boundary condition ensures that waves with the wave number $\kappa$ are completely absorbed while waves with other wave numbers are partly absorbed and partly reflected \cite{Tumulka2022Distribution,Fevens1999Absorbing}. Note that, the Hille-Yosida theorem implies that the Schr\"{o}dinger equation with the above boundary condition has a unique solution for every initial wave function defined on one side of the boundary. 
 
The boundary condition \eqref{ABC_eq}, implies that Bohmian trajectories can cross the boundary $\mathbb{S}$ only outwards and so there are no multi-crossing trajectories. In the Bohmian picture, a detector clicks when and where the Bohmian particle reaches the detection surface $\mathbb{S}$. In fact, it is a description of a “hard” detector, i.e., one that detects a particle immediately when it arrives at the surface $\mathbb{S}$. Nonetheless, it should be noted that the boundary absorbs the particle but not completely the wave. The wave packet moving towards the detector may not be entirely absorbed, but rather partially reflected \cite{Tumulka2022Distribution}.
 
The application of the absorbing boundary condition in arrival time problem was first proposed by Werner \cite{werner1987arrival}, and recently it is re-derived and generalized by other authors using various methods \cite{Tumulka2022Distribution,tumulka2022detection,Dubey2021Quantum,tumulka2022absorbing}. Especially, it is recently shown that in a suitable (non-obvious) limit, the imaginary potential approach yields the distribution of
detection time and position in agreement with the absorbing boundary rule \cite{tumulka2022absorbing}. Moreover, Dubey, Bernardin, and Dhar \cite{Dubey2021Quantum} have shown that the ABR can be obtained in a limit similar but not identical to that considered in the quantum Zeno effect, involving repeated quantum measurements. Recently the natural extension of the absorbing boundary rule to the $n$-particle case is discussed by Tumulka \cite{tumulka2022detection}. The key element of this extension is that, upon a detection event, the wave function gets collapsed by inserting the detected position, at the time of detection, into the wave function, thus yielding a wave function of $(n-1)$ particles. We use this formalism for the investigation of detector back-effect in our double-double-slit setup. In this regard, the corresponding Bohmian trajectories and arrival time distributions are presented in Fig.\,\ref{abctraj}.

\begin{figure}[h!]
	\centering
	\begin{tikzpicture} 
		\draw (0,0) node[above right]{\includegraphics[width=1\linewidth, trim={0.5cm 0cm 0cm 0cm}]{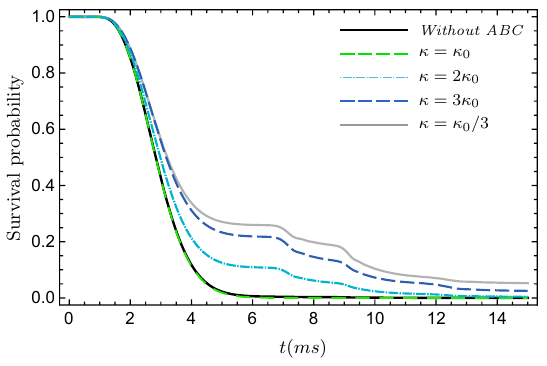}};
	\end{tikzpicture}
	\caption{Survival probability of helium atoms in the two double-slit setup, with different detector characterizing constants  $\kappa$. The atoms are subject to gravity and are absorbed by the detectors with an absorbing boundary. The setup parameters are the same as in Fig.\! \ref{abctraj}.}\label{absorption}
\end{figure}

In our experimental setup, due to the influence of gravity, the reflected portions of the wave packets return to the detector screen, while some of them are absorbed and some are reflected again. This cycle of absorption and reflection is repeated continuously. The associated survival probabilities are plotted in Fig.\,\ref{absorption} for some values of detector parameter, $\kappa=\kappa_0,\, 2\kappa_0,\, 3\kappa_0,\, \kappa_0/3$, where the $\kappa_0$ is defined using classical estimation of particles momentum at the screen as $\kappa_0=m \sqrt{2gY_R}/\hbar$. As one can see in Figs.\,\ref{abctraj} and \ref{absorption}, when $\kappa=\kappa_0$, most of the trajectories are absorbed, and approximately none of them are reflected, which is similar to the case when the detector back effect is ignored. These results show that, at least for the chosen parameters, when we use a proper detector with $\kappa=\kappa_0$, the ideal arrival time distribution computed in the previous section, without considering the detector back effect, produces acceptable results. However, in general, Figs.\,\ref{abctraj} and \ref{absorption} show that the detector back effect cannot be ignored and it leads to new phenomena: i.e., a ``fractal" in the interference pattern. 

\section{Comparition with semiclassical analysis}\label{semiclassical-section}
\begin{figure}[H]
	\centering
	\begin{tikzpicture} 
		\draw (0,0) node[above right]{\includegraphics[width=1 \linewidth, trim={0.5cm 0cm 0cm 0cm}]{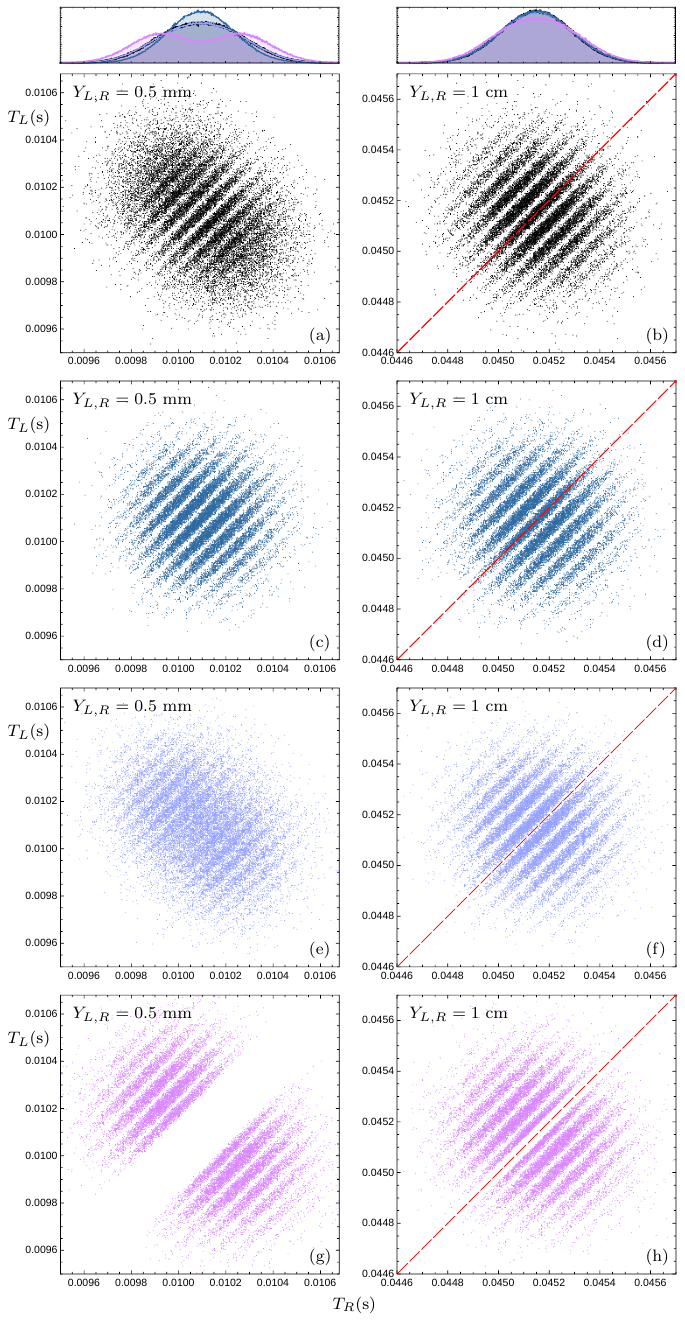}};
	\end{tikzpicture}
	\caption{Comparison of the semiclassical and Bohmian Joint arrival times distributions for sodium atoms, for different  screen positions. The right and left columns correspond to the the screen positions $Y_{L,R}= 1$ cm ( far-field) and $Y_{L,R}= 0.5$ mm (near-field), respectively. Panels (a) and (b) are scatter plots of the Bohmian joint spatio-temporal distribution with considering the wave function collapse effect. Panels (c) and (d) show the semiclassical joint distribution that calculated from classical free fall and quantum initial momentum distribution, $f_P$, with considering the initial positions of all the particles at the $y=0$. Panels (e) and (f) are the semiclassical joint distributions that obtained from the initial phase-space distributions given in Eq.\,\eqref{prod-distibution}. Panels (g) and (h) are the semiclassical joint distributions that obtained from the initial phase-space distributions given in Eq.\,\eqref{corrilated-initial}. Up panels represent the corresponding marginal arrival time distributions.}\label{semiclassical}
\end{figure}

Despite the absence of an agreed-upon fundamental approach for arrival time computation, a semiclassical analysis is routinely used to analyze observed data. This approach is often sufficient, especially when particle detection is done in the far-field regime \cite{vona2013does,Note2}. In this approach, it is assumed that particles move along classical trajectories, and the arrival time distribution is computed using the quantum initial momentum distribution \cite{Das2019Exotic,Shucker1980, wolf2000ion}. It is important to compare the semiclassical analysis with our Bohmian result. To this end, we need to extend the semiclassical approximation  for multi-particles systems in the presence of gravity, which is done as follows:

Using the classical trajectory of a free falling particle, the arrival time is given by
$$
T(y_0,p_{y_0};Y)=p_{y_0}/|m \bm g|+\sqrt{(p_{y_0}/|m \bm g|)^2+2|Y-y_0|/|\bm g|}
$$
where $y_0$ and $p_{y_0}$ are initial particle position and momentum in $y$-direction, respectively, and $Y$ is the position of the horizontal screen. Therefore, the joint semiclassical arrival time distribution is given by
\begin{eqnarray}
\Pi(t_L,t_R)&=&\int f(y^{L}_0,y^{R}_0,p^{L}_{y_0},p^{R}_{y_0})\delta \left(t_L-T(y^{L}_0,p^{L}_{y_0};Y_L)\right) \nonumber\\
&\times & \delta \left(t_R-T(y^{R}_0,p^{R}_{y_0};Y_R)\right)\, dy^{L}_0\, dy^{R}_0\, dp^{L}_{y_0}\,dp^{R}_{y_0} \nonumber
\end{eqnarray}
where 
$f(y^{L}_0,y^{R}_0,p^{L}_{y_0},p^{R}_{y_0})$ is joint initial phase-space distribution, in $y$-direction. However, in standard quantum mechanics, the joint phase-space distribution in a given direction is not a well defined concept. Nonetheless, it is routinely expected that in the far-field regime, the arrival time distribution is independent of initial position distribution and can be calculated just by momentum distribution---In fact, this conjecture is established  for one-particle systems, in some arrival time approaches \cite{das2021times,vona2013does}. In this regard, at first we consider the initial position of all of the particles at $y=0$, and  the resulted semiclassical arrival time distributions are represented in panel (e) and (f) of Fig. \ref{semiclassical}, for near- and far-field regime, respectively. As expected, although in the far-filed regime this approximation more or less is in agreement with Bohmian ones, in near filed semiclassical joint and marginal arrival distributions are very different from their Bohmian counterparts. To avoid this issue, as a more accurate approximation, by ignoring initial position-momentum correlation in the $y$-direction, one may suggest the following initial phase-space distribution: 
\begin{equation}\label{prod-distibution}
f(y^{L}_0,y^{R}_0,p^{L}_{y_0},p^{R}_{y_0})=f_Y(y^{L}_0,y^{R}_0)f_P(p^{L}_{y_0},p^{R}_{y_0}),
\end{equation}
where $f_Y$ and $f_P$ are the initial position and momentum distribution of left-right particles pairs in y-direction, respectively, which can be computed from initial wave function as follow 
$$
f_Y(y^{L}_0,y^{R}_0)=\int |\Psi_0(x_1,y_1,x_2,y_2)|^2\ \xi_Y\ dx_1dx_2
$$
$$
f_P(p^{L}_{y_0},p^{R}_{y_0})=\int |\tilde\Psi_0(x_1,p_{1y},x_2,p_{2y})|^2\ \xi_P\ dx_1dx_2
$$
where
$$
\xi_Y=\theta(x_1)\delta(y^{L}_0-y_1)\delta(y^{R}_0-y_2)+\theta(-x_1)\delta(y^{L}_0-y_2)\delta(y^{R}_0-y_1),
$$
$$
\xi_P=\theta(x_1)\delta(p^{L}_{y_0}-p_1)\delta(p^{R}_{y_0}-p_2)+\theta(-x_1)\delta(p^{L}_{y_0}-p_2)\delta(p^{R}_{y_0}-p_1),
$$
\noindent and $\tilde\Psi_0(x_1,p_{1y},x_2,p_{2y})$ is the Fourier transformation of the initial wave function, $\Psi_0(x_1,y_1,x_2,y_2)$, with respect to $y_1$ and $y_2$ variables. Note that this joint phase-space distribution leads exactly to the quantum initial position and momentum marginal distributions. In the panels (e) and (f) of Fig.\,\ref{semiclassical}, the semiclassical distribution resulting from the phase-space distribution \eqref{prod-distibution}, are plotted in near- and far-field regime, respectively. As a surprising fact, although in the near-field regime the results of this semiclassical analysis are more similar to the Bohmian ones, in the far-field regime this semiclassical arrival time distribution deviates significantly from the Bohmian result: The central interference fringe of the Bohmian joint distribution does not exist in this semiclassical approximation (see red dashed lines in panels (b) and (f)).

In fact, there are various correlated initial phase-space distributions which are consistent with quantum initial position and momentum marginal distributions, however, they lead to different joint arrival time distributions. Merely as an example, see panels (g) and (h) of Fig.\,\ref{semiclassical}, which are generated from another initial phase-space distribution defined as
\begin{equation}\label{corrilated-initial}
\int|\Psi_0(\bm{r}'_1,\bm{r}'_2)|^2 \prod_{i=1,2} \delta(\bm{r}_i-\bm{r}'_i)\delta(\bm{p}-\bm{p}^{\infty}_i(\bm{r}'_1,\bm{r}'_2))d^3r'_i,
\end{equation}
where, $\bm{p}^{\infty}_2(\bm{r}'_1,\bm{r}'_2)$ and $\bm{p}^{\infty}_2(\bm{r}'_1,\bm{r}'_2)$ are the asymptotic momenta of two \textit{free} Bohmian particle with initial positions $(\bm{r}'_1,\bm{r}'_2)$ and initial wave function $\Psi_0$. It is easy to show that above phase-space distribution is consistent with the quantum initial position and momentum distributions \cite{holland1995quantum,durr2010quantum}.

Note that, even in the far-field regime, although  all semiclassical \textit{marginal} arrival time distributions are more or less in agreement with Bohmain results, but the \textit{joint} semiclassical distributions are very sensitive to assumed initial phase-space distributions. These facts suggest that the multi-particle joint arrival time distributions, for example in our suggested double-double-slit setup, can be used to probe the Bohmian arrival time prediction, which is more sensitive than the previously proposed one-particle experiments \cite{rafsanjani2023can,roncallo2023does,das2022double}. Furthermore, it is important to remark that this deviation from semiclassical analysis is predicted without the presence of the challenging effects that have typically been suggested to distinguish arrival time proposals, namely the back-flow effect \cite{bracken1994probability,trillo2023quantum}, multi-crossing Bohmian trajectories \cite{Das2019Arrival,rafsanjani2023can}, and the detector back-effect---as discussed in the previous section. 

\section{Summary and outlook}\label{Summary}
In this work, we have proposed a double-double-slit setup to observe non-local interference in the arrival time of entangled particle pairs. Our numerical study shows a complementarity between one-particle visibility and two-particle visibility in the arrival time interference pattern, which is very similar to the complementarity observed for the arrival position interference pattern \cite{Georgiev2021}. Moreover, our results indicate that the two-particle interference visibility in the arrival time distribution can serve as an entanglement witness, thereby suggesting the potential use of temporal observables for tasks related to quantum information processing \cite{bulla2023nonlocal,anastopoulos2017time}. 

As noted in the introduction, the theoretical analysis of the proposed experiment is more complex than that of a typical double-slit experiment due to several connected fundamental problems, including the arrival time problem and the  detector back-effect problem. We use a Bohmian treatment to circumvent these problems. This approach can be used for a more accurate investigation of various experiments beyond the double-double-slit experiment, such as atomic ghost imaging \cite{khakimov2016ghost},  interferometric gravitometry \cite{geiger2018proposal}, atomic Hong–Ou–Mandel experiments \cite{lopes2015atomic}, and so on \cite{tenart2021observation,brown2023time}, which are usually analyzed in a semiclassical approximation. In such situations, the semiclassical analysis may not lead to a unique and unambiguous prediction, as discussed in section \ref{semiclassical-section}.  

It is worth noting that, based on other interpretations of quantum theory, there are other non-equivalent approaches that, in principle, can be used to investigate the proposed experiment \cite{rafsanjani2023can,roncallo2023does}. However, these approaches need to be extended for entangled particle systems first. Comparing the results obtained by these various approaches can be used to test the foundations of quantum theory. Specifically, it appears that measuring the arrival time correlations in entangled particle systems can sharply distinguish between different approaches to the arrival time problem \cite{anastopoulos2017time}. A more detailed investigation of this subject is left for future works.

\appendix
\section{Estimation of the interference visibility}\label{AppendixB}
A complementarity relationship between one-particle and two-particle interference pattern, $V^2+W^2\leq 1$, is disused in the literature \cite{jaeger1993complementarity,jaeger1995two,Georgiev2021}. The One-particle interference visibility $V$ can be obtained as 
\begin{equation}\label{vis1}
V = (I_{max}-I_{min})/(I_{max} + I_{min}),
\end{equation}
where $I_{max}$ and $I_{min}$ represent the zero-order maximum intensity and first-order minimum intensity, respectively
\cite{xiao2019observing}. In the case of two-particle interference, there are some complexities in the definition of the two-particle interference visibility, and in fact there is no agreed upon definition for general cases. However, some proposal have taken place \cite{luis2003visibility,Georgiev2021}. In the present work, we have used the definition given in Ref. \cite{Georgiev2021}, which is particularly suitable for studying the complementarity relationship in a double-double-slit arrangement. This two-particle visibility definition is briefly described in the following. For a symmetric setup in which the distances between slits are equal for the left and right sides, the arrival time distribution of particles, $\Pi(t_L,t_R)$, exhibit grooves and unit visibility in two diagonal directions aligned with the $t^{\pm}=(t_L\pm t_R)/\sqrt{2}$ and $t_R$ axes in the case of separable states. The rotated marginal distributions projected on  $t_{\pm}$ read as
$$
\Pi^{\pm}(t')=\int \Pi(t_L,t_R) \delta(t'-t^{\pm})\ dt_R dt_L
$$
Using visibilities of the above marginal distributions, $V^{\pm}$, which can be estimated as same as one-particle visibility via Eq.\,\eqref{vis1}, the two-particle visibility is given by \cite{Georgiev2021}:  
\begin{equation}\label{vis2}
W=|V^{+}-V^{-}|.
\end{equation}
The one- and two-particle visibilities are depicted in Fig.\,\ref{visibility}, for various values of $\eta$. In this figure, the error bars and central values are calculated from the average and standard deviation of a set of 10  arrival time joint distributions that each of them generated from $10^4$ Bohmian trajectories.

\bibliography{bibliography}

\end{document}